\documentclass[final,3p,times,twocolumn]{elsarticle}
 \biboptions{comma,sort&compress}
\usepackage{here}
 \usepackage{graphicx}
  \usepackage{epsfig}
\def\nin{\noindent}
\def\beq{\begin{equation}}
\def\eeq{\end{equation}}
\def\bea{\begin{eqnarray}}
\def\eea{\end{eqnarray}}

\newcommand{\JP}{\rm J/$\psi$ }
\newcommand{\pt}{\ensuremath{p_{\rm T}} }
\newcommand{\pp}{pp }
\newcommand{\s}{$\sqrt{s} = 7\ $}

\journal{Nuc. Phys. (Proc. Suppl.)}

\begin{document}

\begin{frontmatter}

%% Title, authors and addresses

%% use the tnoteref command within \title for footnotes;
%% use the tnotetext command for the associated footnote;
%% use the fnref command within \author or \address for footnotes;
%% use the fntext command for the associated footnote;
%% use the corref command within \author for corresponding author footnotes;
%% use the cortext command for the associated footnote;
%% use the ead command for the email address,
%% and the form \ead[url] for the home page:
%%
%% \title{Title\tnoteref{label1}}
%% \tnotetext[label1]{}
%% \author{Name\corref{cor1}\fnref{label2}}
%% \ead{email address}
%% \ead[url]{home page}
%% \fntext[label2]{}
%% \cortext[cor1]{}
%% \address{Address\fnref{label3}}
%% \fntext[label3]{}

\title{\JP production in \pp collisions and in Pb\--Pb collisions at the LHC with the ALICE experiment}

%% use optional labels to link authors explicitly to addresses:
 \author{L.~Manceau for the ALICE Collaboration}
  \address{INFN - Sezione di Torino, Via P. Guiria 1, I-10125 Torino (Italy)}
%\cortext[cor1]{Speaker}
\ead{lmanceau@to.infn.it}

%\author{}

%\address{}

\begin{abstract}
%% Text of abstract
\noindent
%\begin{linenumbers}
We report the recent measurements of the inclusive \JP production in pp collisions at $\sqrt{s} = 2.76$ TeV, in \pp collisions at \s TeV and in Pb\--Pb collisions at $\sqrt{s_{\mathrm{NN}}} = 2.76$ TeV with the ALICE detector at the LHC. The \JP mesons are detected down to \ensuremath{p_{\rm T} = 0} GeV/c via their di-muon decay channel at forward rapidity ($2.5<y<4$) and via their di-electron decay channel at mid-rapidity ($|y|<0.9$). A special emphasis is put on the nuclear modification factor which is compared to the predictions of current \JP production models.
%\end{linenumbers}
\end{abstract}

\begin{keyword}
 \JP meson production \sep \pp collisions \sep Pb\--Pb collisions \sep LHC \sep ALICE experiment
%% MSC codes here, in the form: \MSC code \sep code
%% or \MSC[2008] code \sep code (2000 is the default)
\end{keyword}

\end{frontmatter}

%%
%% Start line numbering here if you want
%%
% \linenumbers

%% main text
%%%%%%%%%%%%
%\begin{linenumbers}
\section{Introduction}
\nin
The unprecedented collision energy reached at the LHC has opened a new era for the study of deconfined QCD matter expected to be created in the extreme conditions characterising early stages of heavy ion collisions. Heavy flavours (charm and beauty) are abundantly produced at LHC energies. As heavy flavours are expected to be created during the initial hard scattering and to coexist with the surrounding medium due to their long life time, their measurement should bring essential information on deconfined QCD matter ~\cite{Alessandro:2006yt}.

While heavy quark pair production is described perturbatively, heavy quark pair bound states are formed via soft non-perturbative processes~\cite{Brambilla:2010cs,Lansberg:2008gk}. Because of this interplay between the perturbative and non-perturbative aspects, quarkonium production in \pp collisions is a unique and a very important test case for QCD. In Pb\--Pb collisions, different theoretical approaches~\cite{Matsui:1986dk,Xu:1995eb,BraunMunzinger:2000px,Thews:2000rj} propose the quarkonium production as a signature of the QCD deconfined matter and, possibly as a thermometer of the medium. 

ALICE (A Large Ion Collider Experiment)~\cite{KAaJINST} is the LHC experiment dedicated to the study of heavy ion collisions. At forward rapidity ($2.5<y<4$), ALICE is equipped with a spectrometer~\cite{KAaJINST} designed to measure muons. Here, we concentrate on the measurement of \JP mesons which decay to two muons. The spectrometer is composed of a passive hadron absorber made of a combination of low and high atomic number material, a beam shield, a $3$ T m dipole magnet, five tracking stations and finally, two trigger stations located behind a $1.2$ m thick iron wall. At mid-rapidity ($|y|<0.9$), \JP mesons are measured via their di-electron decay channel with the ALICE central barrel~\cite{KAaJINST}. Amongst the six sub-detectors composing the central barrel, the Inner Tracking System (ITS)~\cite{KAaJINST} and a $88\ \mathrm{m^{3}}$ Time Projection Chamber (TPC)~\cite{KAaJINST} allow charged particle tracking in a $0.5$ T field. The ITS is made of three layers of detectors using different silicon technologies. The detector layer the closest to the interaction point allows the reconstruction of the interaction vertex. The TPC is filled with a $\mathrm{Ne-CO_{2}}$ gas mixture and is divided in two drift regions by the central electrode. In Pb\--Pb collisions, events are selected according to their degree of centrality by means of the VZERO hodoscope~\cite{KAaJINST} made of two scintillator arrays covering the pseudorapidity range $2.8<\eta<5.1$ and $-3.7<\eta<-1.7$.  %amplitude which is fitted using the Glauber model~\cite{Collaboration:2011rta} to determine intervals in percentages of the nuclear cross section.
The VZERO used together with the ITS also allows to define the Minimum-Bias (MB) interaction trigger.

\section{Data analysis}
\label{Anal}
%While in \pp collisions at \s TeV the results from the analysis of 2010 data are presented, results in \pp collisions at $\sqrt{s} = 2.76$ TeV rely on the analysis of 2011 data. In Pb\--Pb collisions at $\sqrt{s_{\mathrm{NN}}} = 2.76$ TeV, the focus is made on the forward rapidity analysis of 2011 data while results at mid-rapidity were obtain from the analysis of 2010 data. It worth noting that during the Pb\--Pb data taking campaign of 2010, the instantaneous luminosity of collisions was about $30$ times lower than in 2011.
While the analysis at mid-rapidity relies on MB events, the analysis at forward rapidity in \pp (Pb\--Pb) collisions rely on a sample of MB events enriched with single muons (di-muons) by means of the trigger system of the spectrometer.  
In Tab.~\ref{tab:Lint} the integrated luminosity of the analysed data samples are given for the considered systems, energies and rapidity ranges.

\begin{center}
{\scriptsize
\begin{table}[hbt]
 \caption{\scriptsize    Integrated luminosity of data samples used for the analysis.}
   {\small %\scriptsize%\small
\begin{tabular}{lll}
&\\
\hline
\hline
\\
\ &$2.5<y<4$&$|y|<0.9$\\
\\
\hline
\hline
\\
\pp $\sqrt{s} = 2.76$ TeV&$19.9$ nb$^{\rm{-1}}$&$1.1$ nb$^{\rm{-1}}$\\
\pp \s TeV&$15.6$ nb$^{\rm{-1}}$&$5.6$ nb$^{\rm{-1}}$\\
Pb\--Pb $\sqrt{s_{\mathrm{NN}}} = 2.76$ TeV&$70$ $\rm\mu$b$^{\rm{-1}}$&$1.7$ $\rm\mu$b$^{\rm{-1}}$\\
\hline
\hline
\end{tabular}
}
\label{tab:Lint}
\end{table}
}
\end{center}

At forward rapidity, reconstructed particles are identified as muons if their tracks in the tracking stations match with tracks in the trigger stations. At mid-rapidity, the measurement of energy loss provides good discrimination of electrons from the larger yield of hadron.

The electrons and muons passing the analysis cuts~\cite{Aamodt:2011gj,PbPbJP2mumu} are combined in opposite-sign pairs thus allowing the signal extraction from invariant mass distributions. In the di-muon channel the signal shape is described by a Crystal Ball function~\cite{CrystallBall} while the background is parameterized using the sum of two exponentials~\cite{Aamodt:2011gj,PbPbJP2mumu}. In the di-electron channel the signal is obtained by subtracting the background which is estimated using the like-sign pairs invariant mass distribution~\cite{Aamodt:2011gj} or track rotation technique~\cite{Abelev:2012rz}.

The measured \JP ($N_{{\rm J/}\psi}$) yield is normalized to the number of events ($N_{\rm events}$) and corrected for the kinematic acceptance, the reconstruction efficiency and the trigger efficiency ($A\epsilon$) of the detector and for the branching ratio ($\rm BR$) of the di-muon or di-electron decay channel. The inclusive \JP yield is then given by:

\bea
Y_{{\rm J/}\psi} &=& \frac{N_{{\rm J/}\psi}}{N_{\rm events}\ A\epsilon\ BR_{{\rm J/}\psi\rightarrow\mu^{+}\mu^{-}(e^{+}e^{-})}}.
\label{Yield}
\eea

The acceptance and efficiency correction is performed by means of a Monte-Carlo based procedure. A large sample of \JP is generated with PYTHIA~\cite{Skands:2010ak} and is embedded in real events. All the particles are transported in a realistic ALICE detector set up modelled with GEANT3~\cite{GEANT}. An alternative procedure involving background events generated with PYTHIA~\cite{Skands:2010ak} or HIJING~\cite{Deng:2010mv} instead of real events is also used. In \pp collisions, the inclusive \JP yield is converted into cross section using the minimum bias \pp cross section~\cite{Aamodt:2011gj}. The \JP \pp cross section and the inclusive \JP yield from Pb\--Pb collisions are used to calculate the nuclear modification factor, as explain in Sect.~\ref{RAA}. %In Pb\--Pb collisions, the \JP \pp cross section is converted into nuclear modification factor as outlined in .

\nin
\section{Results in pp collisions}
\label{ppxs}
\nin

\begin{figure}[hbt] 
\centerline{\includegraphics[width=7.cm]{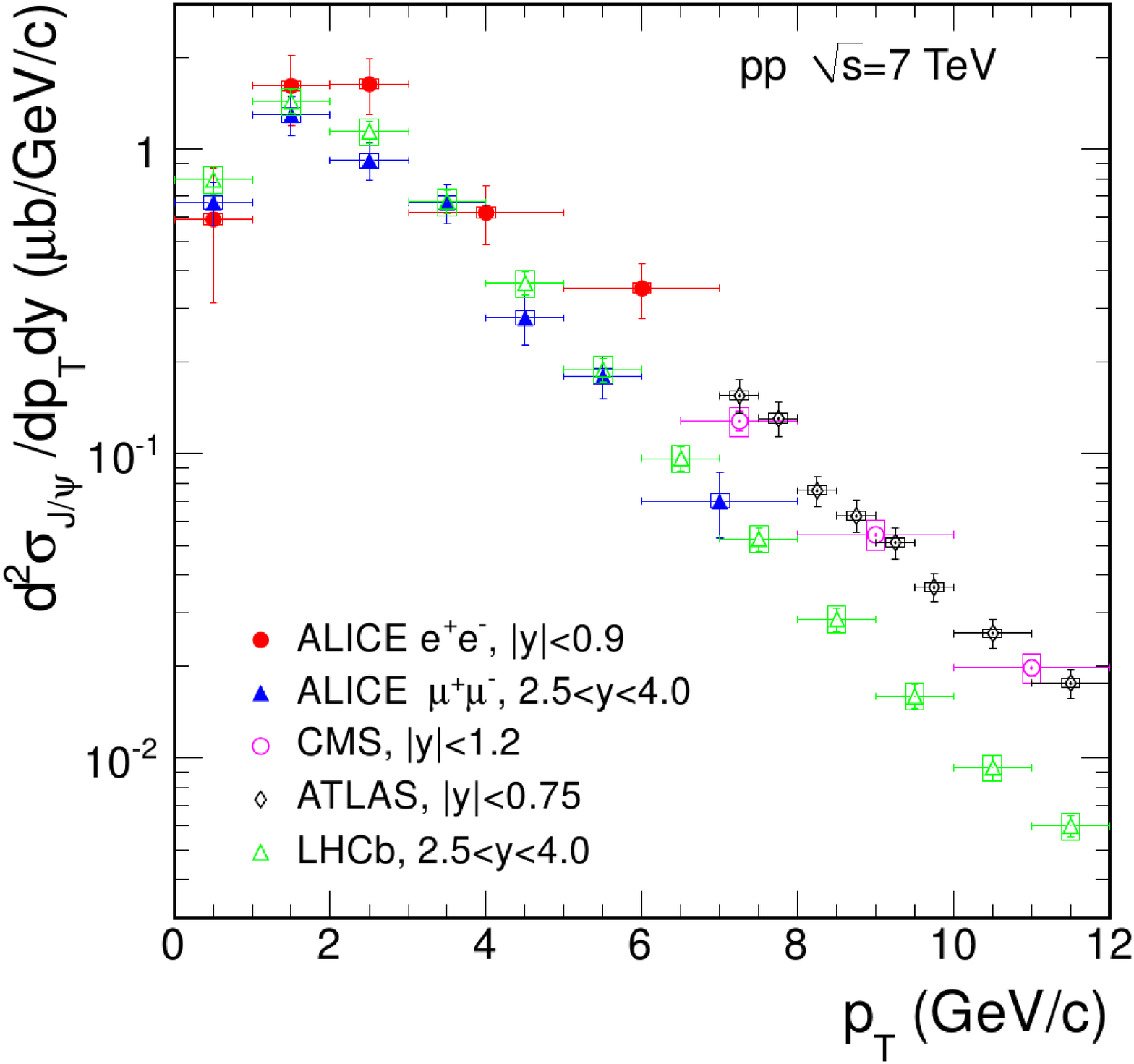}}
%{\epsfig{figure=mpsi2mc.eps,height=70mm}}
\caption{\scriptsize \ensuremath{p_{\rm T}}-differential cross section of inclusive \JP measured at forward and mid-rapidity  in \pp collisions at \s TeV ~\cite{Aamodt:2011gj}. ALICE data are compared with the results from CMS~\cite{Khachatryan:2010yr}, ATLAS~\cite{Aad:2011sp} and LHCb~\cite{Aaij:2011jh}.}
\label{fig0} 
\end{figure}

The \ensuremath{p_{\rm T}}-differential cross sections of inclusive \JP measured in \pp collisions at \s TeV at forward and mid-rapidity~\cite{Aamodt:2011gj} are shown on Fig.~\ref{fig0} together with the CMS~\cite{Khachatryan:2010yr} and ATLAS~\cite{Aad:2011sp} results at mid-rapidity and the LHCb~\cite{Aaij:2011jh} results at forward rapidity. At mid-rapidity, the \pt coverage of the ALICE results is complementary to that from CMS and ATLAS. At forward-rapidity the ALICE results are in agreement with those of LHCb.

\begin{figure}[hbt] 
\centerline{\includegraphics[width=7.cm]{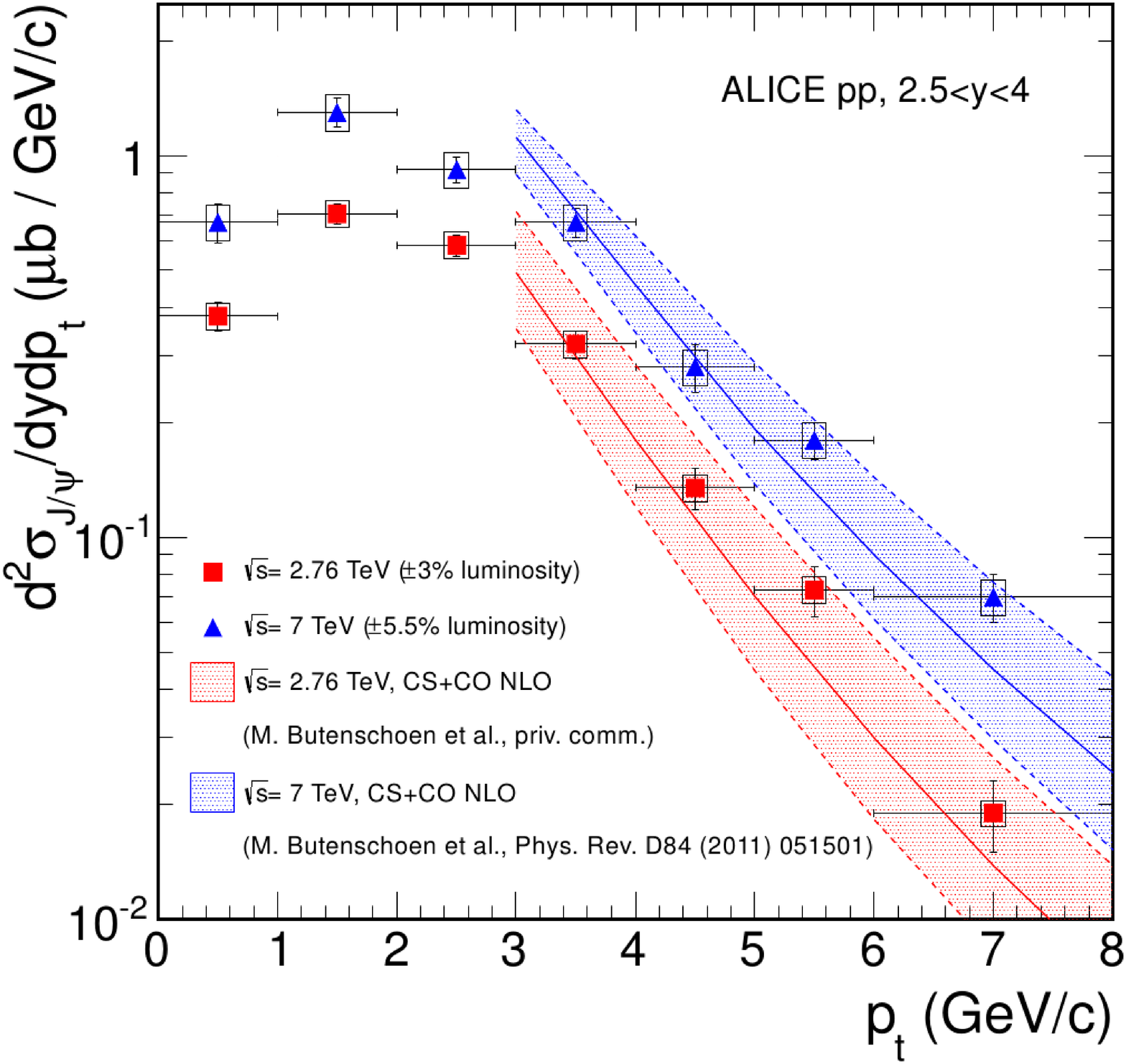}}
\centerline{\includegraphics[width=7.cm]{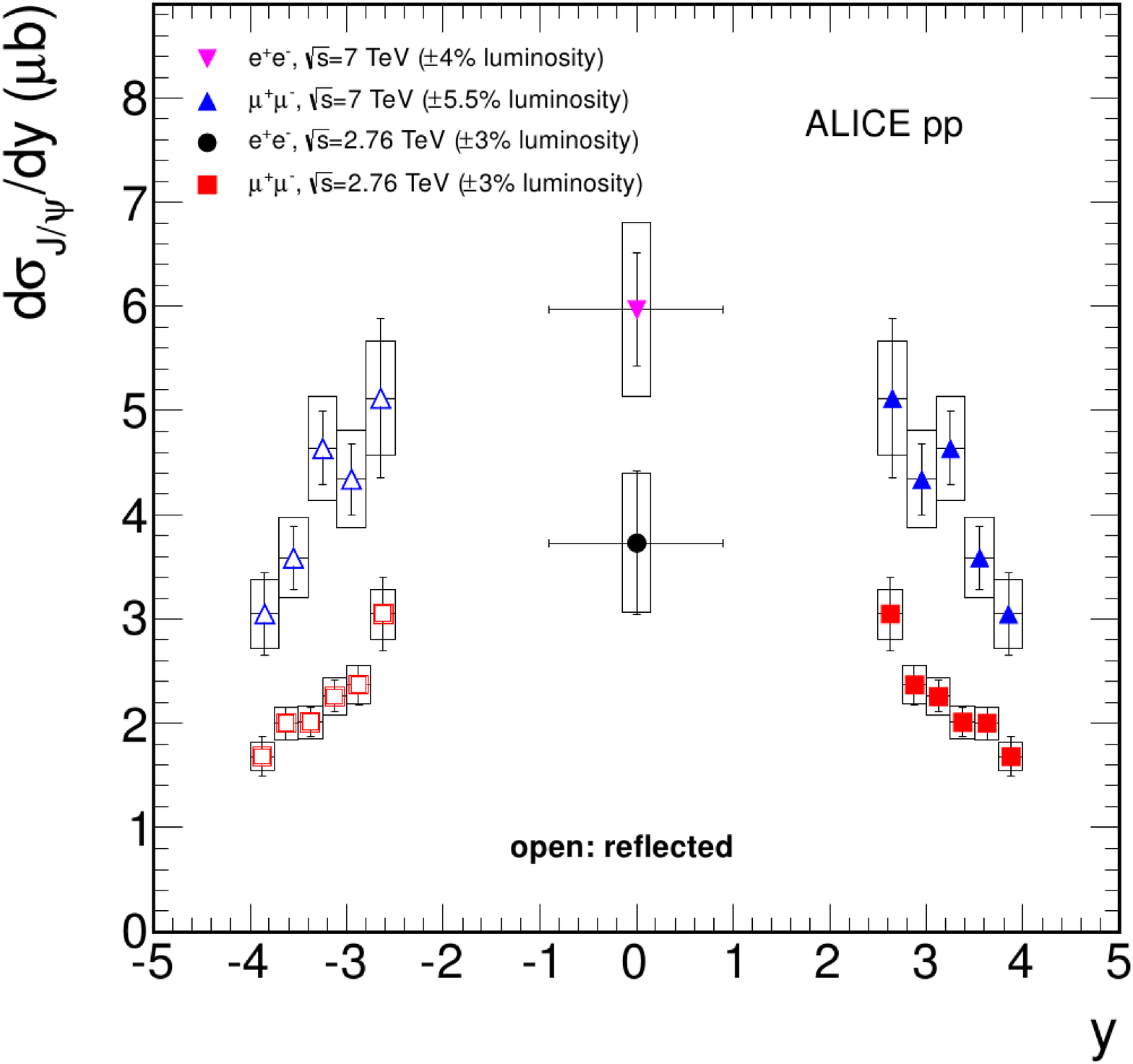}}
%{\epsfig{figure=mpsi2mc.eps,height=70mm}}
\caption{\scriptsize Top: \ensuremath{p_{\rm T}}-differential cross sections of inclusive \JP measured at forward rapidity in \pp collisions at $\sqrt{s} = 2.76$ TeV and at \s TeV~\cite{pp276} compared to NRQCD calculations~\cite{Butenschoen:2011yh}. Bottom: $y$-differential cross sections of inclusive \JP measured at forward and mid-rapidity in \pp collisions at $\sqrt{s} = 2.76$ TeV and at \s TeV~\cite{pp276}}
\label{fig1} 
\end{figure}

The \ensuremath{p_{\rm T}}-differential cross section of inclusive \JP measured at forward rapidity in \pp collisions at $\sqrt{s} = 2.76$ TeV~\cite{pp276} and in \pp collisions at \s TeV~\cite{pp276} is presented on the top panel of Fig.~\ref{fig1}. The results at both energies are compared with the predictions of a NRQCD calculation~\cite{Butenschoen:2011yh} which includes both color singlet and color octet terms at NLO. The $y$-differential cross section of inclusive \JP measured at forward rapidity in \pp collisions at $\sqrt{s} = 2.76$ TeV~\cite{pp276} and in \pp collisions at \s TeV~\cite{pp276} is presented on the bottom panel of Fig.~\ref{fig1}. The contribution of \JP from B-hadron feed-down in \pp collisions at \s TeV was measured at mid-rapidity and account for $10-15\%$ of the total inclusive yield~\cite{BJP}.

\nin
\section{Results in Pb\--Pb collisions}
\label{RAA}
\nin

The \JP yield ($Y_{{\rm J/}\psi}$) measured in Pb\--Pb collisions in different centrality classes is combined with the inclusive \JP cross section measured in pp collisions at the same energy (see Sect.~\ref{ppxs}) to form the nuclear modification factor ($R_{\rm AA}$) defined
as:
\bea
%R_{\rm AA}^{\rm i}=\frac{Y^{\rm i}_{{\rm J/}\psi}}{\langle T_{\rm AA}^{\rm i}\rangle\sigma_{{\rm J/}\psi}^{\rm pp}}.
R_{\rm AA}=\frac{Y_{{\rm J/}\psi}}{\langle T_{\rm AA}\rangle\sigma_{{\rm J/}\psi}^{\rm pp}}.
\eea
A Glauber model calculation~\cite{Aamodt:2010cz} is used to determine the average nuclear overlap function ($\langle T_{\rm AA}\rangle$) and the average number of participating nucleons ($\langle  N_{\rm part}\rangle$) for each centrality class used in this analysis. %The average number of participating nucleons is used to described the dependence of the nuclear modification factor on the size of the fireball.

\begin{figure}[hbt] 
\centerline{\includegraphics[width=8.cm]{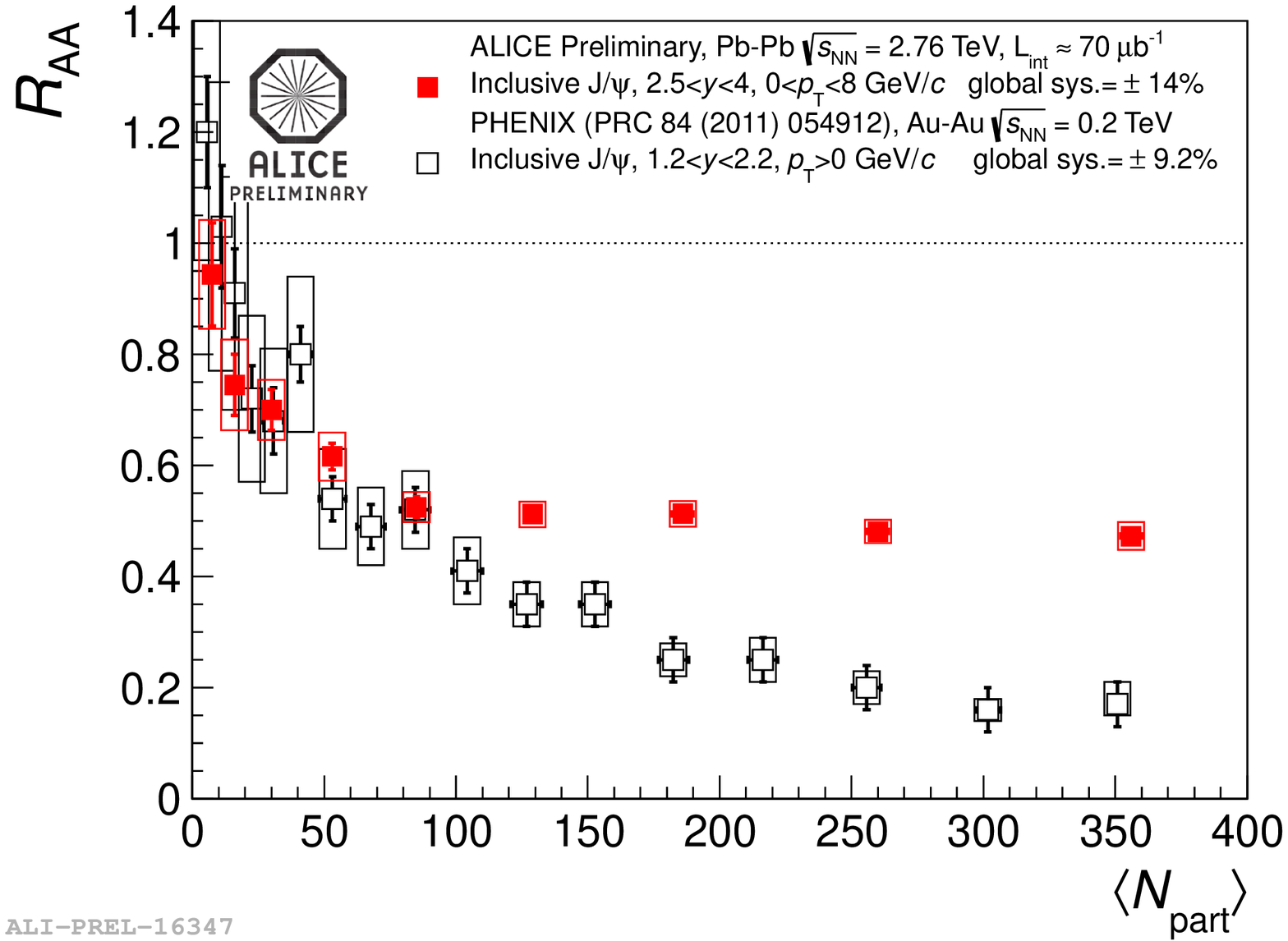}}
\centerline{\includegraphics[width=8.cm]{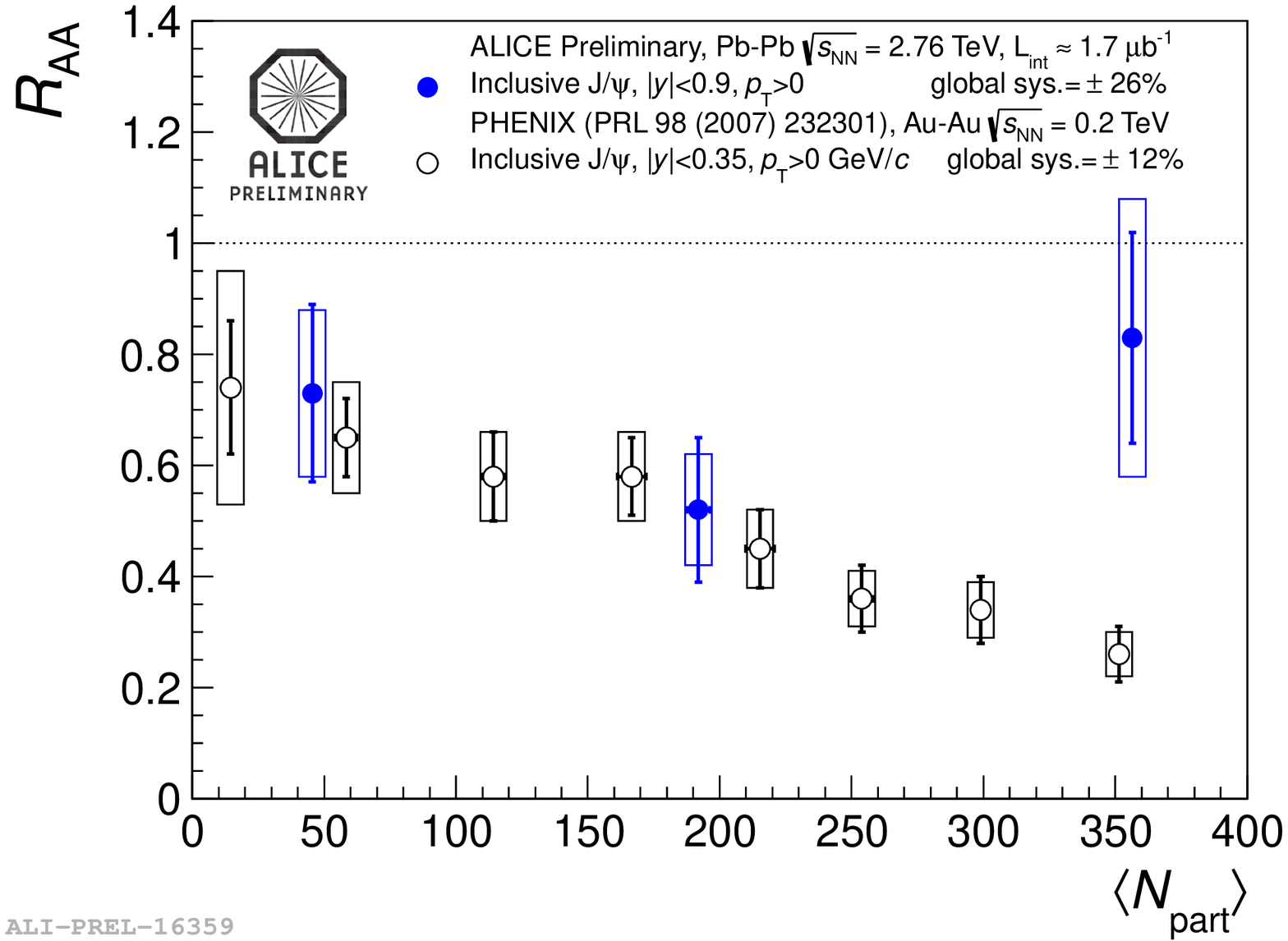}}
%{\epsfig{figure=mpsi2mc.eps,height=70mm}}
\caption{\scriptsize Inclusive \JP nuclear modification factor measured at forward rapidity (top) and mid-rapidity (bottom) compared to PHENIX results.}
\label{fig2} 
\end{figure}

The inclusive \JP $R_{\rm AA}$ measured by ALICE at $\sqrt{s_{\mathrm{NN}}} = 2.76$ TeV in the range $2.5 < y < 4$ ($|y|<0.9$) and $\ensuremath{p_{\rm T}}>0$ GeV/c is shown as a function of $\langle N_{\rm part}\rangle$ in the top (bottom) panel of Fig.~\ref{fig2}. The comparison with the PHENIX measurements at $\sqrt{s_{\mathrm{NN}}} = 200$ GeV at forward rapidity~\cite{Adare:2011yf} and mid-rapidity~\cite{Adare:2006ns} shows that the ALICE inclusive \JP $R_{\rm AA}$ is almost a factor of three larger for $\langle N_{\rm part}\rangle>180$ ($\langle N_{\rm part}\rangle = 350$) at forward rapidity (mid-rapidity). In addition, ALICE results do not exhibit a significant centrality dependence for $\langle N_{\rm part}\rangle>50$.

\begin{figure}[hbt] 
\centerline{\includegraphics[width=8.cm]{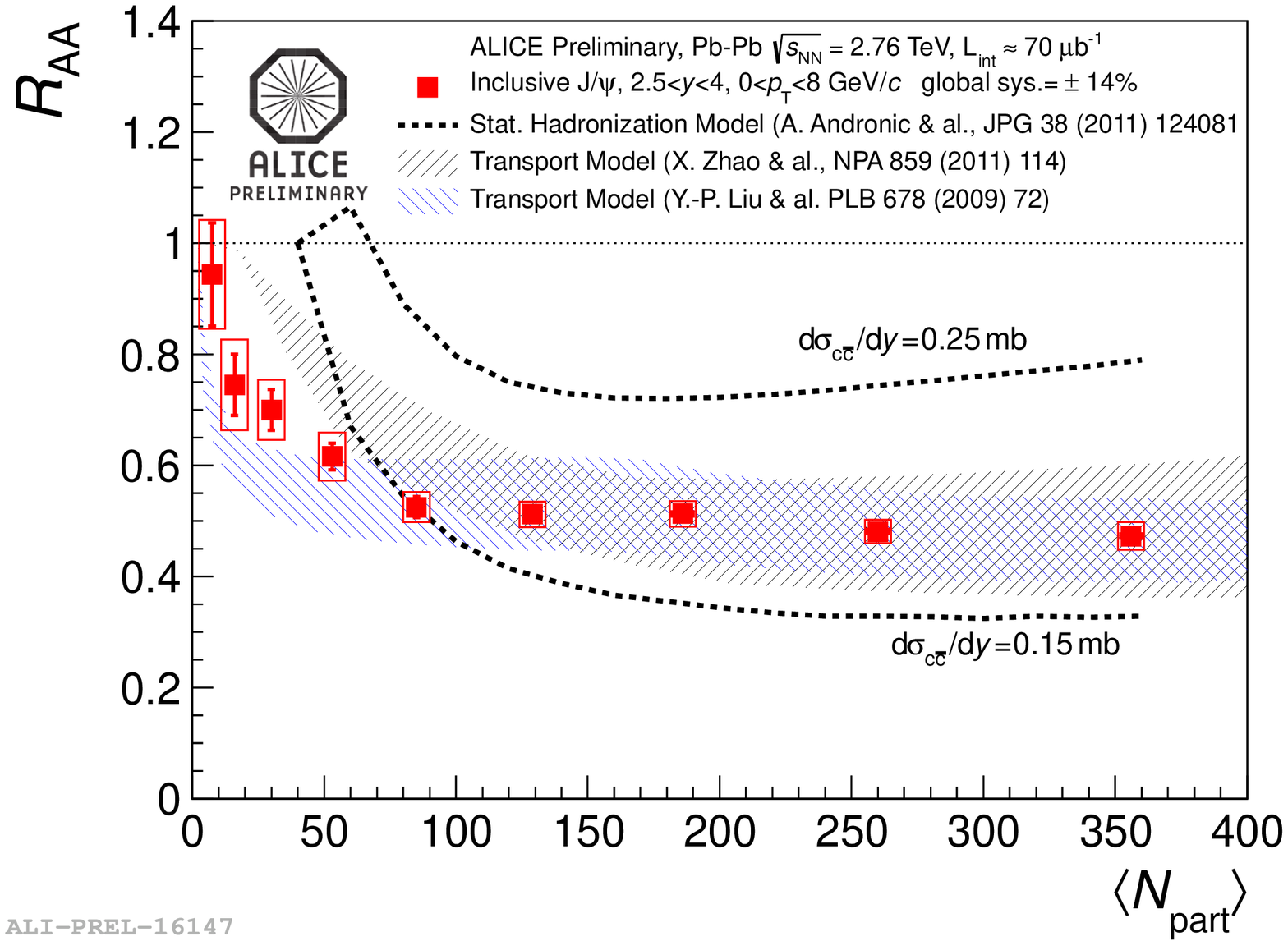}}
\centerline{\includegraphics[width=8.cm]{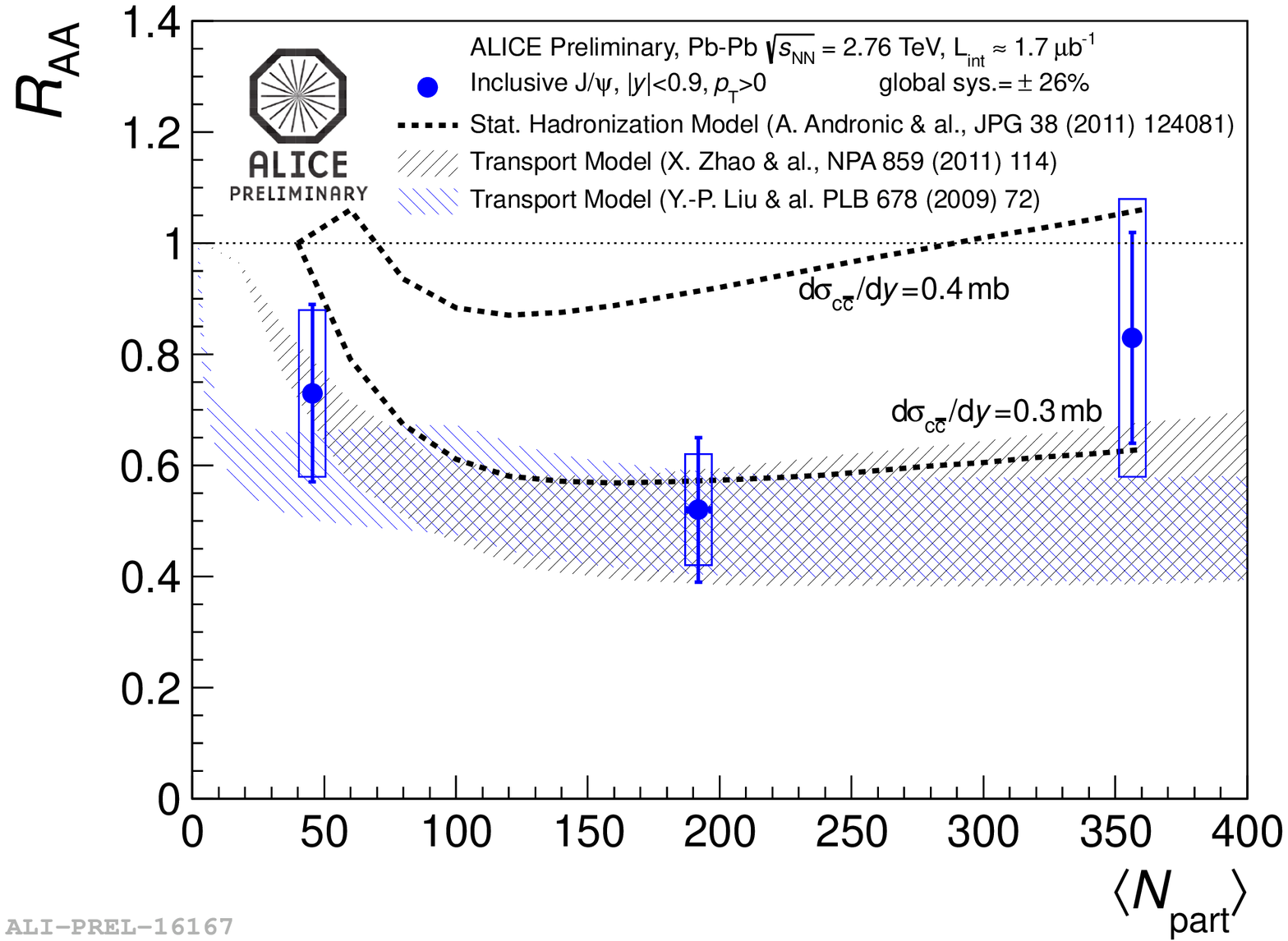}}
%{\epsfig{figure=mpsi2mc.eps,height=70mm}}
\caption{\scriptsize Inclusive \JP nuclear modification factor measured at forward rapidity (top) and mid-rapidity (bottom) compared to the prediction of models including a \JP (re)generation component in the deconfined medium.}
\label{fig3} 
\end{figure}

In Fig.~\ref{fig3} ALICE measurements are compared with theoretical models that include a \JP (re)generation component from deconfined charm quarks in the medium. The Statistical Hadronization Model~\cite{Thermo} assumes deconfinement and a thermal equilibration of the bulk of the ${\rm c\bar{c}}$ pairs. Then charmonium production occurs at phase boundary by statistical hadronization of charm quarks. The prediction is given for two values of ${\rm d}\sigma_{\rm c\bar{c}}/{\rm d}y$, since it has not yet been measured for Pb\--Pb collisions. The two transport model results~\cite{Zhao:2011cv,Liu:2009nb} presented in the same figure differ mostly in the rate equation controlling the \JP dissociation and regeneration. Both are shown as a band which connects the results obtained with (lower limit) and without (higher limit) shadowing. The width of the band can be interpreted as the uncertainty of the prediction. In both transport models, the amount of regenerated \JP in the most central collisions contributes about $50\%$ of the measured yield, the rest being from initial production.

\nin
\section{Conclusion}
Inclusive \JP production was measured in \pp collisions at $\sqrt{s} = 2.76$ TeV and at \s TeV. The \ensuremath{p_{\rm T}}-differential cross section of inclusive \JP measured at forward rapidity in \pp collisions at \s TeV is reproduced by NRQCD calculations. The nuclear modification factor of inclusive \JP in Pb\--Pb collisions at $\sqrt{s_{\mathrm{NN}}} = 2.76$ TeV was measured both at forward and mid-rapidity. The \JP suppression was found to be smaller than at RHIC and does not exhibit significant centrality dependence. Models including full or partial production of \JP from charm quarks in the deconfined medium reproduce the data.
%\end{linenumbers}
%\label{}
\nin

\end{document}